\documentclass[aps,prd,twoside,twocolumn,floatfix,letterpaper]{revtex4-1}
\usepackage[english]{babel}
\usepackage[latin1]{inputenc}
\usepackage[dvipdfm]{graphicx}
\usepackage[dvipdfm]{rotating}
\usepackage{graphicx,pdfpages}
\usepackage{subfig}
\usepackage{float}
\usepackage{lipsum,fancyhdr}
\usepackage{textcomp}
\usepackage{subfig}
\usepackage{array}
\usepackage{amsmath,amssymb}
\usepackage[T1]{fontenc}
\usepackage{mathptmx}

\fancypagestyle{plain}{%
\chead{}

}

\newcommand{\hcond}{\usefont{T1}{phv}{mc}{n}} 
\makeatletter
\def\section{\@startsection{section}{1}{0pt}{-3.5ex plus -1ex minus -.2ex}{2.3ex plus .2ex}{\large \hcond}}
\def\subsection{\@startsection{subsection}{2}{\z@}{-3.25ex plus -1ex minus -.2ex}{1.5ex plus .2ex}{\hcond}}

\begin{document}

\title{The polar ring galaxy AM\,2040-620 and its possible companion}
\author{Priscila Freitas-Lemes$^{1}$, Irapuan Rodrigues$^{1}$ and Maximiliano
 Fa{\'u}ndez-Abans$^{2}$}\email[]{priscila@univap.br, irapuan@univap.br and max@lna.br}

\affiliation{$^{1}$ Universidade do Vale do Para{\'i}ba, Av. Shishima Hifumi 2911, 
Cep 12244-000, S{\~a}o Jos{\'e} dos Campos, SP, Brazil $^{2}$.
MCT- Laborat{\'o}rio Nacional de Astrof{\'i}sica, Caixa Postal 21, 
CEP:37.504-364, Itajub{\'a}, MG, Brazil.}

\begin{abstract}
Polar ring galaxies (PRGs) are peculiar systems where a gas-rich, nearly polar ring
surrounds a host galaxy. They are the result of galaxy interactions that form mainly 
by tidal accretion of material from a gas rich donor galaxy. There is a 
number of formation mechanisms for PRGs: minor or major mergers, tidal accretion events,
or direct cold gas accretion from filaments of the cosmic web. These 
objects can be used to probe the three-dimensional shape of dark matter haloes, provided
that the ring is in equilibrium with the gravitational potential of the host galaxy. 
The polar ring galaxy, AM\,2040-620, which has not yet been well studied, is the subject 
of this work. This galaxy contains an almost perpendicular warped ring and one possible 
companion galaxy to the NW. 
The radial velocity of this object is 3301$\pm$65 Km s$^{-1}$ 
and is part of a group of fifteen possible polar ring galaxies, according to the literature. 
In order to better understand this system, images and long slit spectra were observed with 
the 1.60 m OPD/LNA telescope.
In the I band image, the outer parts of the ring are not symmetrical. A disturbance in 
the Eastern side and a faint plume were detected. Two small satellites are located to 
the north. The bulge is elliptical but not perfectly symmetrical in this image. The B-band
image shows material that extends beyond the ring in the western and eastern directions.
After processing, the B-image shows that the possible companion galaxy 
2MASX J20441668-6158092 has a tidally disturbed disk. Its radial velocity is unknown, 
but the spectroscopy, which is still under analysis, will furnish this information. 
\end{abstract}

\maketitle\thispagestyle{plain}

\section{Introduction}
Polar ring galaxies (PRGs) constitute a rare class of interacting systems and consist
of an early-type, lenticular, elliptical or even spiral host galaxy,
surrounded by a ring of gas, dust and young stars orbiting in a nearly polar plane 
(e.g. \cite{1990AJ....100.1489W}, \cite{1993A&A...268..103A}, \cite{1997yCat..41290357F}, 
\cite{2001MNRAS.322..689R}, \cite{2006A&A...446..447R}). 

In these objects, the velocities of the ring and the host galaxy are quite close, with
an almost balanced state. The ring material appears to be in regular rotation about 
the galaxy center (e.g. \cite{2004AJ....128.2013C}, \cite{1987ApJ...314..457V}, 
\cite{1997AJ....113..585A}) and is presumably stabilized in some way. According to 
Iodice (2002) \cite{2002AJ....123..195I}, for a PRG to become stable, the formation of the
ring must follow a certain logic, in which the ring size is 
strongly related to the amount of matter in the host galaxy (visible + dark). 

The presence of two almost perpendicular angular momentum vectors cannot be explained 
through the collapse of a single protogalactic cloud; therefore a ``second event'' must 
have occurred in the formation history of these objects. However, no one knows for sure 
how these objects are formed. There are three formation mechanisms for PRGs 
\cite{2003A&A...401..817B}:
minor or major mergers, tidal accretion events or direct cold gas accretion from cosmic 
web filaments. These objects can be used to probe the three-dimensional 
shape of dark matter haloes, provided that the ring is in equilibrium with the 
gravitational potential of the host galaxy. 

In this paper, we study the PRG AM\,2040-620. The galaxy has an elliptical bulge. 
Its with radial velocity is 3301$\pm$65 km s$^{-1}$ (this paper), very similar (within the errors) to the value measured from HI data by van Driel (2002) \cite{2002A&A...386..140V} (3335$\pm$24 km s$^{-1}$). The ring is not radially thick and presents a warped appearence.

 
\begin{figure}
 \centering
  \includegraphics[width=5cm]{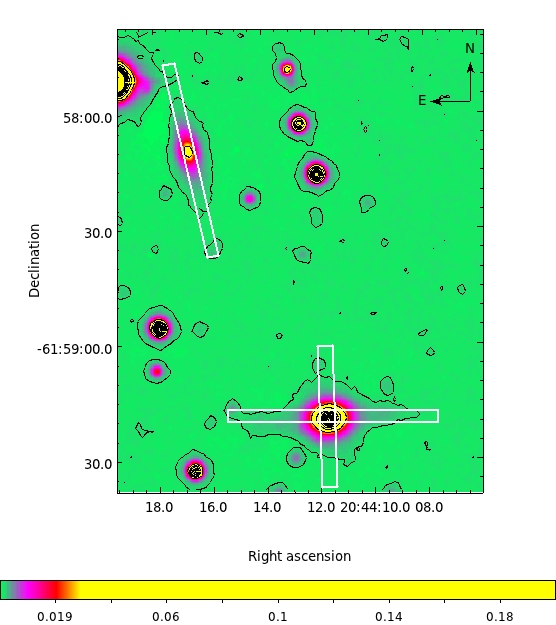}
  \caption{B-band image of NGC\,5122 with slit position.}\label{inicial} 
\end{figure}

\section{Observations and reductions}

Broad band imagery was carried out with the 1.6-m telescope at the OPD (MCT/LNA), Brazil. 
The telescope was equipped with direct imaging camera$\sharp$1 
with CCD$\sharp$106 (1024x1024 square pixels). The data were acquired with standard Johnson 
U, B, V, R and I filters. 


Long slit spectroscopy was done a with the same telescope equiped with a Cassegrain 
spectrograph and CCD$\sharp$105 (2048x2048 square pixels). During the observations, 
we used grating of 600 lines mm$^{-1}$, centered at 5800\AA. We took 
three slit positions (see Figure \ref{inicial}): (1) along the ring (0\textdegree); (2)along the galaxy's major axis 
(host) (88\textdegree); and (3) along the companion galaxy's major axis (77\textdegree). 
CCD data reduction was performed in the standard manner using IRAF (\textit{Image Reduction 
and Analysis Facility}) package. HeAr lamps were measured before and after each exposure in 
order to provide accurate wavelength calibration. The spectra were flux calibrated using
 standard stars observed at similar airmasses. 

\section{Results}

\begin{figure}
  \includegraphics[width=6cm]{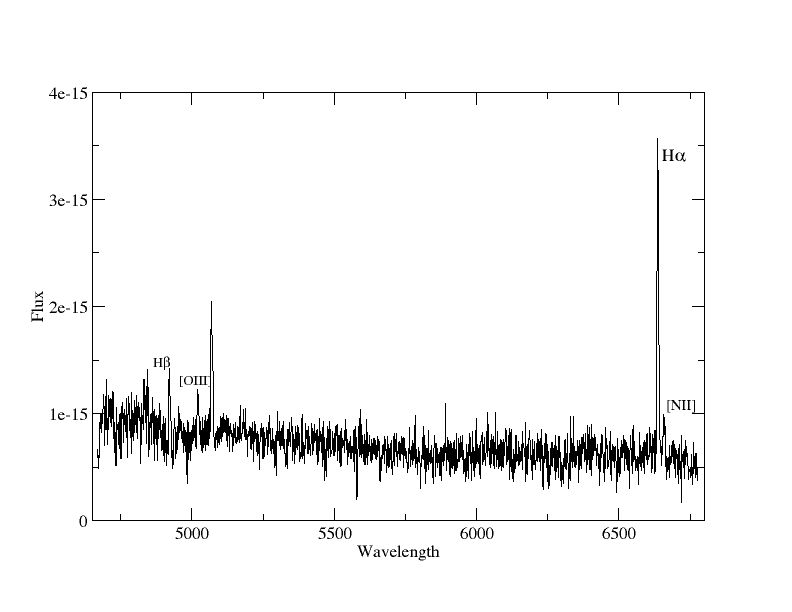}
  \caption{Companion Galaxy Spectra.}\label{espectro} 
\end{figure}

The host galaxy of this system is surrounded by a nearly polar ring (inclined 88\textdegree).
The eastern region of the system has some traces of perturbation and tide. 
R-band image shows an elliptical bulge, which is not completely 
symmetrical, with some strains principally in NE and SW. There are two small satellites to 
the SE. We measured its radial velocity as 3301$\pm$65 km s$^{-1}$, value close
to the literature
\cite{2002A&A...386..140V}.

The ring has a wavy behavior. At the extremes of the ring is low surface
brightness leftover material. This material may be a remnant of the 
interaction process that originated the ring.

\begin{figure}
  \centering
  \includegraphics[width=5cm]{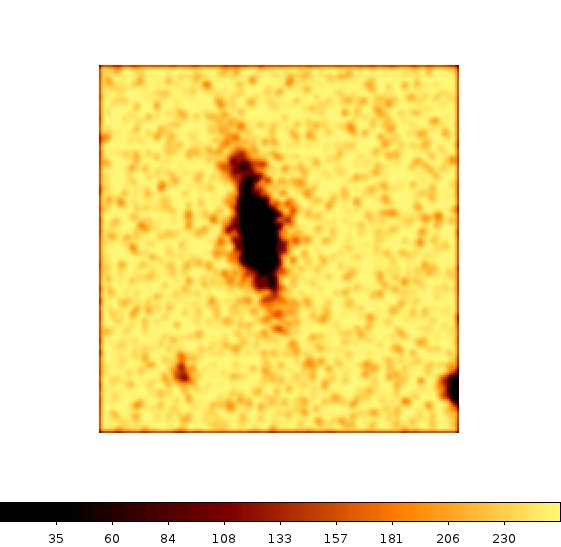}
  \caption{B-band image of companion galaxy. With possible tidal disturbances.}\label{compa} 
\end{figure}

Figure \ref{espectro} shows the spectrum of the companion galaxy of the system, 
where the main lines of the system are marked.
Many PRGs have companions, because they are often the donors 
of material for the formation of the ring. AM\,2040-620 is a galaxy in the NW, 
which had hitherto unknown speed. After our observations, we measure its radial velocity 
as 3331$\pm$ 47 Km s$^{-1}$, which is very close to that of the PRG. These
results indicate that these galaxies are possibly part of a group, which went
through a process of interaction that resulted in the formation of the ring.
Image B, processed through a square $50\times50$ pix$^2$ kernel, and 
subtracted from the original image, is shown in Figure \ref{compa}. It
highlights high frequency structures, making more evident the warped outer
parts of the disk, and suggesting a tidal origin. There is no bridge or material
that binds the rest of this interacting galaxy system.

\section{Conclusions}

This work allows us to conclude that this system is a galaxy with polar ring, 
which has a companion galaxy with visible deformities, possibly due to the effects of the 
interaction process. A bridge or tails interconnect the PRG. The 
ring is corrugated and symmetrical. The host galaxy is elliptical and has some deformities.

\section{Acknowledgements}

The author thanks FAPESP for the scholarship granted under process 2010/17136-4.

\bibliography{publicacao.bib}
\bibliographystyle{unsrt}

\end{document}